# What makes a good concept anyway ?

Naren Khatwani
*Department of Data Science*
New *Jersey Institute of Technology*
Newark, NJ, USA
nk88@njit.edu

James Geller
*Department of Data Science*
New *Jersey Institute of Technology*
Newark, NJ, USA
geller@njit.edu

**A good medical ontology is expected to cover its domain completely and correctly. On the other hand, large ontologies are hard to build, hard to understand, and hard to maintain. Thus, adding new concepts (often multi-word concepts) to an existing ontology must be done judiciously. Only "good" concepts should be added; however, it is difficult to define what makes a concept good. In this research, we propose a metric to measure the goodness of a concept. We identified factors that appear to influence goodness judgments of medical experts and combined them into a single metric. These factors include concept name length (in words), concept occurrence frequency in the medical literature, and syntactic categories of component words. As an added factor we used the simplicity of a term after mapping it into a specific foreign language. We performed Bayesian optimization of factor weights to achieve maximum agreement between the metric and three medical experts. The results showed that our metric had a 50.67% overall agreement with the experts, as measured by Krippendorff's alpha.**

## I. INTRODUCTION

One definition of a (medical) ontology is … "a structured framework that represents knowledge within the medical domain. It consists of a set of concepts and the relationships between them, facilitating the sharing and integration of medical knowledge across different systems and applications" [1, 2].

The Stanford BioPortal [3] is a repository of over 1,130 medical ontologies and terminologies. Small ontologies are of limited utility, but large ontologies are hard to build, hard to understand, and hard to maintain. Maintainers of widely used medical ontologies often refuse to add new concepts to keep the maintainability manageable, unless there is a "business use case" for a new concept. This was specifically observed for the National Cancer Institute thesaurus (NCIt) [4].

In medical ontology development and knowledge representation, ensuring the quality and reliability of concepts is paramount. Intuitively, only "good" concepts should be added to an existing ontology. This paper presents a novel approach to evaluating the "goodness" of concepts for medical ontologies. The goal of this research is to develop a robust, metric that can systematically assess and rank concepts based on several factors, ultimately leading to more efficient and effective ontology extension. Our objectives include identifying relevant factors, developing a numeric formula for computing the goodness of a concept, and evaluating the metric based on a comparison with medical experts.

We utilized the Unified Medical Language System (UMLS) as a source of concepts for developing the metric. The UMLS is a comprehensive collection of biomedical and health vocabularies that facilitates interoperability between computer systems within the healthcare domain [5].

This paper details the methodology used to develop the concept goodness formula, including the criteria for selecting factors and integrating them into a cohesive metric. We also present the results of empirical tests to validate the formula, demonstrating its effectiveness in distinguishing high-quality concepts from lower-quality ones.

The significance of this work lies in its potential to standardize concept evaluation in medical ontologies, providing a tool for researchers and practitioners to ensure the integrity and usability of their knowledge representations. By improving the assessment of concept goodness, this project makes it possible to enhance the quality and reliability of medical ontologies, ultimately supporting better decision-making and knowledge dissemination in the medical field.

## II. BACKGROUND

The question of what a concept is in philosophy has a history of at least 2,300 years. According to Salmieri [6], Aristotle is "generally held" to define a concept as a "putative grasp of a mind-independent universal object that exists somehow in or derivatively on the many particular objects to which the concept applies." Concepts are the building blocks of thoughts and understanding, allowing us to categorize and make sense of the world around us. They are essential for communication, reasoning, and the development of knowledge across various domains [6].

In Medical Informatics, a more practical notion of a computational *concept* is widely used. For simplicity we will refer to it just as "concept." In Gruber's view [1] a concept "is a fundamental building block of an ontology. … Concepts … represent classes or categories of objects…"

The minimum requirement for a (computational) concept is that it is unique and distinguishable from all other concepts in a system. This can be achieved by assigning each concept a unique ID number. It is assumed that this concept represents an object in the real world, however, this object may be abstract, e.g., "health." The power of concepts is derived from the way how they are combined with each other in an ontology by links that reflect relationships in the real world.

The comprehension of ontologies is greatly helped by a visual analogy. Thus, a concept is a "box" in a diagram, and a relationship is an arrow connecting two boxes (limiting ourselves to binary relationships). Typically, additional "local" information may be associated with each concept. The core element of local information is its "preferred term" (using SNOMED meta-terminology) [7] which is the way how humans would likely communicate about this concept. If a concept can be referred to by several terms, those are considered synonyms of the preferred name.

As concepts describe classes (see above) there is a strong assumption that there are instances of a concept. Whether these instances are part of an ontology, or of a knowledge graph, or of a database is a subject of discussion [8]. Requirements for the properties of medical ontologies have been described in a widely-cited and subsequently revised paper of Jim Cimino [8, 9]. However, the question of what makes a concept "good" is not deeply explored.

## III. METHODS

In this study, a "good" concept is defined as one that meets several key criteria, ensuring its utility, clarity, relevance, and significance within the medical professional context. Our working hypothesis of what makes a good concept is as follows. A good concept should be easily understood without causing confusion or requiring extra effort, facilitating effective communication and support knowledge transfer. For example, the term "cholesterol" is likely good for most people as it is straightforward and widely understood. In contrast, "posttranslational protein modification site" might be considered bad as it can be confusing and requires specialized knowledge. It needs to be asserted that what is a good concept for a medical practitioner may not be a good concept for a layperson. For example, a layperson would probably prefer "heart attack" over "myocardial infarction," while an MD would not.

Furthermore, a good medical concept might be commonly heard or used in recent public conversations or media, reflecting its relevance and familiarity among the general population. "Opioid Crisis" is likely good, as it is a widely recognized and relevant concept in public health discussions. On the other hand, "trepanation" ("drilling a hole") may be considered bad, due to its decreasing presence in modern medical discourse and its lack of relevance to current medical practices.

A good concept is also significant in describing many entities in the real world, encompassing both tangible and abstract entities, demonstrating its broad applicability and importance in various domains. The concept of "metabolism" is likely good, because it applies to various real-world contexts, including health, nutrition, and fitness, whereas "leech" may be considered bad due to its limited relevance and application in today's medical world. (Thus, what is "good" changes slowly over time.)

Lastly, a good concept can make a difference in a legal context, where its precise definition could impact legal outcomes. This underscores the necessity for concepts to be clearly defined and universally understood to ensure their effectiveness in critical and high-stakes situations. Thus, the decision when death has occurred is highly significant in law. Therefore, death is a "good" concept from an ontology point of view.

In this paper, we will use several factors to evaluate the goodness of concepts. These factors include Brevity, Frequency of Occurrence, German Language Mappability, Presence in a Medical Dictionary, and Grammatical Structure. (The choice of German will be explained below.) By integrating these factors into a cohesive evaluation metric, we aim to systematically assess and rank the quality of medical concepts, thereby enhancing the reliability and usability of medical ontologies.

We will now explain the choice of German as a mapping language. The UMLS provides multi-lingual support, and we observed that the same concept is often expressed with fewer (or more) words in a different language. Subscribing to the "efficiency of language" theory, if a language creates a single word for a concept, this concept must be important to its speakers [10, 11]. Thus, we noted that in many cases German uses a single word for concepts that are described in English (in the UMLS) by more than one word, because German speakers tend to construct multi-noun words. Thus, we added the "mappability" into a single word in a foreign language as a goodness criterion for a concept expressed in English with a multi-word term. This choice was bolstered by the fact that one of the authors is fluent in German.

## IV. RELATED WORK

The development and maintenance of medical ontologies are crucial for organizing and standardizing knowledge across various biomedical domains. While there are existing methodologies for cross-referencing terms across multiple ontologies, these approaches often focus on broader associations rather than evaluating individual concept quality [12]. Our work emphasizes ontology enrichment at the concept level, where each concept is rigorously tested for its "goodness" before being integrated into an ontology as a new concept. This approach ensures that only high-quality concepts, those that serve unique purposes and provide meaningful contributions, are added to the ontology.

Existing research highlights how online information sources like Wikipedia can be utilized to retrieve medical-related information and enrich ontologies effectively. However, several key factors affect the proper utilization of these resources. First, Wikipedia and other online forums have an open editing policy, which limits the ability to assess contributors' medical proficiency. This leads to concerns about inaccuracies and the risk of misinformation [13, 14]. Additionally, while community extensibility is significant for developing ontologies, research has highlighted explicitly that it presents significant challenges. Ontologies often begin with a small group of contributors, making initial development straightforward, but as the group widens, reaching a consensus on adding new concepts becomes increasingly difficult [15]. To address these challenges, we propose evaluating concepts using additional factors that account for a medical concept's online presence, conciseness, and occurrence in research publications.

This approach recognizes that while there are numerous medical concepts available, it is only feasible to include judiciously chosen exemplars in ontologies. Adding concepts based on subjective judgment can lead to redundancies and compromise the ontology's effectiveness. This is particularly important in recently developed ontologies such as the SOHO Social Determinants of Health (SDoH) ontology [16]. Their focus is to avoid redundant concepts from the outset by implementing rigorous criteria for concept inclusion. This proactive approach ensures that each concept added serves a unique purpose and contributes to the overall integrity and utility of the ontology [17].

## V. DATA COLLECTION AND PREPROCESSING

To construct a set of concepts to be used for evaluating the quality of the goodness metric developed in this paper, we turned to the largest source of medical concepts, the Unified Medical Language System (UMLS) [18] as the primary data repository. However, the concepts to be added to SDoH and other ontologies in the future will come from varied information sources, most of them English text, and the UMLS is not intended to be the primary source of concepts.

The data collection and preprocessing stages were crucial for ensuring the quality and relevance of the concepts included in our analysis. The UMLS provides comprehensive details about medical concepts and their associated semantic types, including concept names, Concept Unique Identifiers (CUIs), and the semantic types from the UMLS Semantic Network linked to each CUI, offering a detailed classification system. This facilitates precise and consistent identification of medical concepts across diverse healthcare applications, enhancing interoperability and data integration.

To ensure the dataset's relevance to our research on biomedical concepts, we applied semantic filtering to exclude certain semantic types that were not pertinent to our focus. Specifically, we excluded certain semantic types (Table 1).

TABLE I. EXCLUDED SEMANTIC TYPES

| Semantic Types |
|---|
| Organic Chemical |
| Pharmacologic Substance |
| Biologically Active Substance |
| Antibiotic |
| Chemical |
| Chemical Viewed Functionally |
| Clinical Drug |
| Amphibian |
| Animal |
| Bird |
| Fish |
| Mammal |
| Plant |
| Reptile |
| Medical Device |
| Research Device |
| Element |
| Ion |
| Isotope |
| Gene or Genome |
| Population Group |
| Eukaryote |

These exclusions were based on the understanding that these semantic types are not central to medical concepts. "Clinical drugs" is the semantic type of many complex names such as 1,3-dimethylxanthine, which we maintain are unnatural even to medical professionals. This step was necessary to eliminate chemical compounds and prescription drugs that do not align with our research focus on (bio)medical concepts.

Further, we filtered the dataset to retain only English terms. This ensured that our analysis would be conducted on a linguistically consistent dataset. After applying these preprocessing steps, we had a refined dataset containing approximately 3 million concept names.

VI. METHODOLOGY

Having completed the preprocessing steps outlined above, the resulting dataset was the basis for further analysis and development of our goodness formula. The following sections describe the key factors we considered: Brevity, Frequency of Occurrence, German Language Mappability, Presence in a Medical Dictionary, and Grammatical Structure.

*A. Brevity*

The word count of a term (concept name) refers to the number of words it contains. Surprisingly, the UMLS [18] contains concept names with more than 80 words. For example, the concept with the CUI C4523185 has the concept name *Endovascular repair of infrarenal aorta by deployment of an aorto-aortic tube endograft including pre-procedure sizing and device selection, all nonselective catheterization(s), all associated radiological supervision and interpretation, all endograft extension(s) placed in the aorta from the level of the renal arteries to the aortic bifurcation, and all angioplasty/stenting performed from the level of the renal arteries to the aortic bifurcation; for rupture including temporary aortic and/or iliac balloon occlusion, when performed (eg, for aneurysm, pseudoaneurysm, dissection, penetrating ulcer, traumatic disruption).* While we are not doubting that this *term* could be useful to medical professionals, we don't see a justification for calling it a single concept.

The longest term we found contains 202 words. Concepts with shorter word counts are generally easier to understand and use, which can enhance their overall goodness. To ensure fairness in comparison and to standardize the impact of word count across all concepts, we normalized the word count values. The normalization was done using the formula:

$$Normalized\ Word\ Count = 1 - \left(\frac{Word\ Count}{Max\ Word\ Count}\right) \quad (1)$$

In (1), the concept with the highest word count receives a normalized value of 0, while the concept with the lowest word count (one word) receives a normalized value very close to 1. This normalization process allows us to incorporate brevity as a factor in our goodness formula, ensuring that concepts with fewer words are rated higher in terms of ease of understanding and use.

*B. Frequency of Occurrence (PubMed)*

The frequency of occurrence of a concept in PubMed is one indicator of how often it appears in the biomedical literature. Higher frequencies suggest that a concept is widely recognized and utilized within the medical community, making its frequency a crucial factor in assessing its relevance and importance.

To determine the frequency of occurrence for each concept, we used the Entrez Programming Utilities (E-utilities), a set of public APIs provided by the National Center for Biotechnology Information (NCBI) [19] to interact with the Entrez system. This approach allowed us to efficiently query PubMed, accessing its vast database of biomedical literature to gather data on how frequently each concept appears in titles, abstracts, and Medical Subject headings (MeSH) terms.

Thus, using the package above, we queried the PubMed database for each concept in our dataset. This method comprehensively assessed each concept's presence in the biomedical literature. Following this, the frequency data were then integrated into our dataset, quantitatively measuring each concept's recognition and usage within the medical community. Note that we counted synonymous terms separately.

*C. German Language Mappability*

As mentioned, any concept expressed by a single word appears to be important to the speakers of its language. As German tends to create long compounded nouns, word length in German is likely to be a better measure of goodness than word length in English. Initially we looked for German concept names in the UMLS itself. However, we did not find enough of those to make mappability to German a useful component of a goodness metric. Therefore, we switched our approach to using language translation software.

To determine the German equivalents of English concept names, we utilized googletrans 3.0.0 [20] – a Python library that interfaces with the Google Translate API – facilitating the comparison and evaluation of concepts across languages. This allowed us to translate the English terms in our dataset into German. Once the translations were obtained, we evaluated the German equivalents based on the length and structure of the terms. Next, we assigned scores based on how well the German equivalents matched the English terms according to their word counts and structures. Additionally, compound words were rewarded with an additional score of 0.1, up to a maximum of 1.0, recognizing German compound words' linguistic nuances and compactness.

For concepts with 1 to 3 words, the corresponding scoring is presented in Table II. For example, the English term "blood pressure test" translates to "Blutdrucktest," which is a German equivalent coming under the category of *same or fewer words*. "Blutdrucktest" is a compound word in the context of the score. Thus, it gets a Final Score of 1.0 with no compound word reward added, because 1.0 is already the maximum.

TABLE II. SCORING FOR CONCEPT NAMES WITH 1 TO 3 WORDS

| Condition | Base Score | Reward | Final Score |
|---|---|---|---|
| German equivalent has the same or fewer words | 1.0 | +0.0 | 1.0 |
| German equivalent has up to +1 word | 0.8 | +0.0 | 0.8 |
| German equivalent has more than +2 words | 0.5 | +0.0 | 0.5 |
| Compound Word reward (if detected) | Varies by condition | +0.1 | Base Score + 0.1 (max 1.0) |

On the other hand, "acute kidney injury" translates to "Akute Verletzung der Niere" in German. Additionally, the translation here does not form a compound word. Thus, the final score assigned to this English term is 0.8.

TABLE III. SCORING FOR CONCEPT NAMES WITH 4 TO 6 WORDS

| Condition | Base Score | Reward | Final Score |
|---|---|---|---|
| German equivalent has fewer words | 1.0 | +0.0 | 1.0 |
| German equivalent has the same number of words | 0.8 | +0.0 | 0.8 |
| German equivalent has more than +1 or +3 words | 0.5 | +0.0 | 0.5 |
| Compound Word reward (if detected) | Varies by condition | +0.1 | Base Score + 0.1 (max 1.0) |

For concepts with 7 to 20 words, the corresponding scoring is presented in Table IV. Due to the length of those concepts, we will not provide examples for them.

TABLE IV. SCORING FOR CONCEPT NAMES WITH 7 TO 20 WORDS

| Condition | Base Score | Reward | Final Score |
|---|---|---|---|
| Translation has 20% fewer words than the original | 0.9 | +0.0 | 0.9 |
| Translation has fewer words than the original | 0.7 | +0.0 | 0.7 |
| Translation has the same or more words than the original | 0.4 | +0.0 | 0.4 |
| Compound Word reward (if detected) | Varies by condition | +0.1 | Base Score + 0.1 (max 1.0) |

For concepts with 21 to 80 words, the corresponding scoring is presented in Table V.

TABLE V. SCORING FOR CONCEPT NAMES WITH 21 TO 80 WORDS

| Condition | Base Score | Reward | Final Score |
|---|---|---|---|
| Translation has 20% fewer words than the original | 0.7 | +0.0 | 0.7 |
| Translation has the same number or more words than the original | 0.4 | +0.0 | 0.4 |
| Compound Word reward (if detected) | Varies by condition | +0.1 | Base Score + 0.1 (max 1.0) |

In all cases, compound words in German translations were rewarded with an additional score of 0.1, acknowledging their linguistic efficiency. This scoring system allows for a refined evaluation of the concept goodness, utilizing the characteristics of the German language.

*D. Dictionary Presence and Grammatical Structure*

The presence of a concept in reputable medical dictionaries is a strong indicator of its validity and acceptance. We checked each concept name against Merriam-Webster's Medical Dictionary [21], to determine its dictionary presence. Concepts included in this dictionary are considered reliable, contributing positively to their goodness score.

In addition to checking dictionary presence, we analyzed concept names using Part-of-Speech (POS) tags to validate meaningful combinations. This involved analyzing the grammatical structure of the concepts and identifying valid combinations. The combinations considered for evaluation included pairs and sets of nouns, adjectives, verbs, and prepositional phrases, each assigned a specific score. Figure 1 elaborates the process that we used. Prompting of ChatGPT was the major tool used to discover the goodness of certain grammatical combinations.

**User:**

> There are various parts of speech combinations possible for medical terms one being noun+noun, give me all the possible combinations you have been trained on.

**ChatGPT:**

When it comes to generating possible parts of speech (POS) combinations for medical terms, here are several combinations that are commonly used in medical terminology:

1. **Noun + Noun** (e.g., "Heart Disease")
2. **Adjective + Noun** (e.g., "Chronic Pain")
3. **Noun + Verb** (e.g., "Disease Progresses")
4. **Noun + Prepositional Phrase** (e.g., "Pain in Chest")
5. **Adverb + Verb** (e.g., "Rapidly Progresses")
6. **Verb + Noun** (e.g., "Treat Disease")
7. **Adjective + Noun + Noun** (e.g., "Acute Respiratory Distress")

*[Result truncated by User]*

**User:**

> Consider the research done on these combinations and noun+noun being the best combination as per the general intuition rank all the other combination on a scale from 0-1 and give examples for each.

**ChatGPT:**

Based on general intuition and research in medical terminology, here's a ranking of different parts of speech (POS) combinations on a scale from 0 to 1. The ranking reflects their relevance, frequency, and utility in medical contexts, with 1 being the best and most relevant, and 0 being the least relevant. I'll also provide examples for each combination:

1. **Noun + Noun** (Rank: **1.00**)
   • **Example**: "Heart Disease"
   • **Rationale**: This combination is highly descriptive and frequently used in medical terminology.
2. **Adjective + Noun** (Rank: **0.90**)
   • **Example**: "Chronic Pain"
   • **Rationale**: This combination is also common, particularly in describing conditions or symptoms.
3. **Noun + Prepositional Phrase** (Rank: **0.85**)
   • **Example**: "Pain in Chest"
   • **Rationale**: This combination is often used in clinical descriptions to specify locations or contexts.
4. **Verb + Noun** (Rank: **0.80**)
   • **Example**: "Treat Disease"
   • **Rationale**: This combination is essential in clinical instructions and treatment protocols.
5. **Adverb + Verb** (Rank: **0.75**)
   • **Example**: "Rapidly Progresses"
   • **Rationale**: Useful in describing the manner or speed of medical conditions, though less common.
6. **Adjective + Noun + Noun** (Rank: **0.70**)
   • **Example**: "Acute Respiratory Distress"
   • **Rationale**: This combination adds more specificity and is often seen in complex medical terms.
7. **Noun + Noun + Noun** (Rank: **0.65**)
   • **Example**: "Blood Pressure Measurement"
   • **Rationale**: Used to describe more complex procedures or concepts, but less frequent than simpler structures.

*[Result truncated by User]*

Fig. 1. Prompting Session with ChatGPT

In creating ontologies, it is important to consider grammatical structures and their implications for semantic understanding [22]. Cognitive research on noun-noun combinations [23, 24] highlights that they align closely with general language processing and comprehension patterns, making them easy to understand and use. In such phrases the second noun is normally the "head noun," while the first noun functions as modifier.

Given this foundation, we prioritized noun-noun combinations, assigning them the highest scores, followed by other combinations in descending order of scores. The scores were generated using GPT- 4. Figure 1 shows the (truncated) interactions with ChatGPT.

This approach of checking dictionary presence and validating POS combinations supported the assessment of the concepts' goodness. By leveraging the dictionary and an advanced language model, we enhanced the overall reliability of the goodness score.

*E. Concept Goodness Formula*

We developed a goodness formula to evaluate and rank concepts systematically based on the factors above. This formula is a weighted average taking into account Brevity, Frequency of Occurrence, German Language Mappability and Dictionary Presence (including Grammatical Structure). The weights between 0 and 100 assigned to each factor reflect their relative importance in determining the overall goodness of a concept.

The Goodness Formula is defined as:

$$GS = \frac{(Br * w1) + (FO * w2) + (GLM * w3) + (DP * w4)}{w1 + w2 + w3 + w4} \quad (2)$$

In (2), Goodness Score (GS) is a weighted mean where Br (Brevity) measures the conciseness of a concept named based on its word count, FO (Frequency of Occurrence) indicates how frequently a concept appears in relevant literature or databases, German Language Mappability (GLM) assesses the ease with which a concept can be linguistically integrated into German, and Dictionary Presence (DP) (including Grammatical Structure) determines whether a concept is included in a recognized medical dictionary. Each weight (*w1, w2, w3, w4*) corresponds to the importance of its corresponding variable in assessing concept quality. This structured approach ensures that our evaluation of concept goodness is balanced, reflecting both quantitative and qualitative aspects of the concepts. One research question is to find the best combination of weights.

VII. EMPIRICAL TESTING FOR DETERMINING WEIGHTS

Initially, we assigned the following weights, based solely on our intuitions (Table VI).

TABLE VI. INITIAL WEIGHT ASSIGNMENT

| Contributing Factors | Weight |
|---|---|
| Brevity | 20 |
| Frequency of Occurence | 25 |
| German Language Mappability | 25 |
| Dictionary Presence and Grammatical Structure | 30 |

## A. Survey Iteration 1

We conducted a survey with 11 participants, including three physicians (Physician 1, 2 and 3) and eight laymen, all of whom had backgrounds in science or general medical fields to evaluate the effectiveness of our concept goodness framework. The evaluation involved 50 concepts: 25 randomly selected from those with the highest goodness scores according to Equation (2) and 25 from the concepts with the lowest scores. Participants rated each concept as "Good," "Moderate," or "Bad."

We used Krippendorff's α (alpha) [25, 26] to assess the interrater reliability among participants, which resulted in a score of 42%, indicating moderate agreement and variability in expert opinions. Further analysis of the physicians' responses revealed a close-to-normal distribution (see Fig. 2) in one physician's ratings, effectively categorizing concepts into three buckets: Good, Moderate, and Bad. This normal distribution highlighted the potential of our scoring system to align with expert evaluations, guiding our decision to focus on the physician's responses that exhibited this pattern.

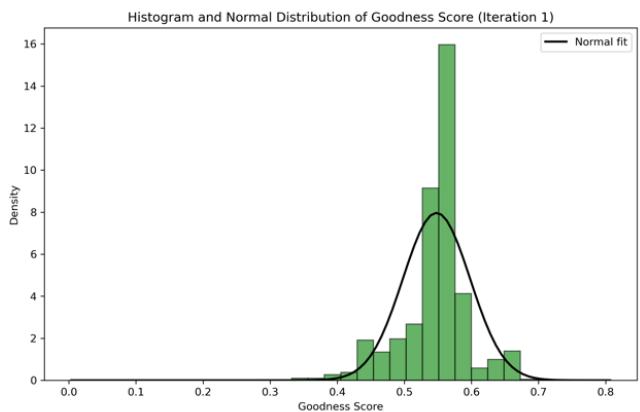

Fig. 2. Normal distribtuion of Physician 3's ratings, categorizing concepts into Good, Moderate and Bad Buckets (Survey Iteration 1)

We next attempted to optimize the weights using Grid Search and Random Search methods. However, we found that the number of possible weight combinations was too large to explore. Grid Search involves evaluating every possible combination of weights within a predefined range, quickly becoming computationally expensive and impractical for large search spaces. Random Search, while less exhaustive, still requires evaluating numerous random combinations, which can be inefficient and may not effectively zero in on the most promising regions of the search space.

Initially, we utilized the grid search algorithm [27] to explore the potential weight combinations for the four factors used in our analysis Brevity, Frequency of Occurrence, German Language Mappability, Presence in a Medical Dictionary, and Grammatical Structure, each varying from 0 to 100 with a step size of 10. However, this approach evaluated the entire search space, resulting in $11^4$ =14,641 possible combinations. Of these combinations, 286 were producing goodness scores matching Physician 3 sufficiently closely. However, there was no systematic or logical way to select a final weight combination from these 286 possibilities.

Additionally, even random search methods [27, 28], while less exhaustive, would still inadvertently focus on a subset of combinations, risking overemphasis of certain factors. These considerations called for a more refined approach, prompting the shift to Bayesian Optimization [29]. This approach uses probabilistic models to guide the search process, focusing on the most promising regions of the weight space and iteratively refining the search, based on previous evaluations. According to "Bayesian optimization replaces the optimization of an expensive-to-evaluate function by a series of optimizations of a cheap criterion." Indeed, Bayesian Optimization proved to be more effective in narrowing down the search space and identifying the most favorable set of weights for our context.

The weights derived by the above computation are shown in Table VII.

TABLE VII. WEIGHTS DERIVED BY PHYSICIAN 3'S RATINGS

| Contributing Factors | Weight |
|---|---|
| Brevity | 49 |
| Frequency of Occurence | 19 |
| German Language Mappability | 10 |
| Dictionary Presence and Grammatical Structure | 20 |

## B. Survey Iteration 2

We revised our dataset's goodness scores by taking the resultant weights derived from the physician 3's survey results in Iteration 1 that were closest to a normal distribution (see Fig. 2) as a baseline. We then assigned those baseline weights in the formula and recalculated the goodness of all the concepts. In this survey, three physicians (Physicians 1, 2, and 3) participated.

Following this we created a similar survey as in the previous iteration involving 50 concepts: 25 randomly selected from those with the highest goodness values and 25 from those with the lowest scores according to our weighted criteria. However, in this survey the set of possible answers was increased. The participants rated each concept as "Definitely a Medical Concept," "Acceptable as a medical concept," "Neutral," "Doubtful as a medical concept," or "Definitely not a medical concept."

These responses were then mapped into three buckets of goodness: "Good," "Moderate," and "Bad." Specifically, "Definitely a Medical Concept" and "Acceptable as a Medical Concept" were mapped to "Good," "Neutral" was mapped to "Moderate," and "Doubtful as a Medical Concept" or "Definitely not a Medical Concept" were mapped to "Bad."

Further analyzing the survey responses, we utilized Krippendorff's α [25, 26] and assessed the interrater reliability among participants, which resulted in a score of 50.67%, indicating moderate agreement and variability in expert opinions. A detailed analysis of the physicians' responses revealed a normal distribution in Physician 1's ratings (Figure 3), effectively categorizing concepts into three buckets: "Good," "Moderate," and Bad. This normal distribution offered improved distinction between goodness buckets compared to the first iteration, leading to a better differentiation of concept goodness values.

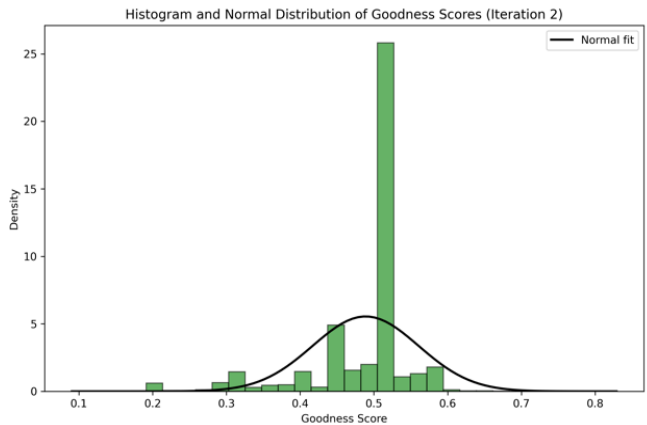

Fig. 3. Normal distribtuion of Physician 1's rating categorizing concepts in Good, Moderate and Bad Buckets (Survey Iteration 2)

Furthermore, we used Bayesian optimization, as in Iteration 1, to find the best matching set of weights. They are given in Table VIII.

TABLE VIII. FINAL WEIGHT ASSIGNMENTS FOR FORMULA (2)

| Contributing Factors | Weight |
|---|---|
| Brevity | 22% |
| Frequency of Occurence | 27% |
| German Language Mappability | 31% |
| Dictionary Presence and Grammatical Structure | 15% |

## VIII. DISCUSSION

There is a growing need for ontology enrichment and extension with the goal of adding non-redundant medical concepts, which expand an ontology's completeness of domain coverage. Having identified gaps in the prior research, where ontologies are evaluated as a whole, we created a metric for concept goodness that evaluates medical concept names for the purpose of including or not including them in an ontology.

### A. Limitations

A limitation of the current work is the small number of healthcare professionals who contributed to it and the limited number of iterations. Conducting more iterations might have given us a better understanding of the weights. Additionally, we have not considered synonymous medical terms in this study. Thus, one concept may be described by more than one term. The choice of German as a mapping language was justified, but limiting. More languages might be useful as mapping targets.

When evaluating a metric that is supposed to reflect human judgment, it is not sufficient to compare the metric with a single expert or even several experts. Rather it is necessary to also compare human experts to each other. If human experts show a wide variability, then this defines a range of what can be expected for the comparison between the metric and the human experts. If humans disagree, then one cannot expect the metric to perfectly agree with any single human expert. Thus, the observed Krippendorff values are not surprising.

## IX. CONCLUSIONS AND FUTURE WORK

In this research, we strived to fill a gap in ontology development and extension, namely the lack of a good metric to judge whether specific concepts should be added to an ontology. For this purpose, we developed a four-part metric that takes into account the factors – Brevity, Frequency of Occurrence, German Language Mappability, Presence in a Medical Dictionary, and Grammatical Structure. Furthermore, we assumed these four components would not be equally impactful and therefore should be combined as a weighted sum.

We performed two iterations of comparing the results of the metric with the results of evaluations by human experts and laymen. In the first iteration, after conducting a survey with medical professionals, we achieved an overall agreement of 42%, as calculated by Krippendorff's α. After modifying the weights and assigning new goodness scores in the second iteration, the physicians' agreement increased to 50.67%.

Up to this point we have evaluated concepts by themselves. However, when a new concept needs to be inserted into an existing ontology, there is also the question to what degree it overlaps with existing concepts. Even if a new concept has a high score according to our goodness metric, if a similar concept exists, the new concepts might be judged as insufficiently different to be also included. In future work we will study this aspect of concept goodness that will incorporate not just individual concepts, but also the target ontology.

## ACKNOWLEDGEMENT

Research reported in this publication was supported by the National Center For Advancing Translational Sciences of the National Institutes of Health under Award Number UM1TR004789. The content is solely the responsibility of the authors and does not necessarily represent the official views of the National Institutes of Health.

The authors would like to thank the following medical professionals for their generous contributions of time and expertise to this research: Dr. Preet Motiramani, M.D in Obstetrics and Gynecology, Dr. Pooja Belligund, MD in Pulmonary Medicine, Dr. Rajesh Koppikar, MDS, Dr. Kamlesh Khatwani, M.B.B.S, Dr. Sunita Prasad, M.D in Pathology and Dr. Nikita Khatwani, M.B.B.S. Their invaluable insights and feedback have significantly enriched the quality of this study.